\documentclass{article}

\usepackage{arxiv}

\usepackage[utf8]{inputenc} 
\usepackage[T1]{fontenc}    
\usepackage{hyperref}       
\usepackage{url}            
\usepackage{booktabs}       
\usepackage{amsfonts}       
\usepackage{nicefrac}       
\usepackage{microtype}      
\usepackage{lipsum}
\usepackage{graphicx}
\usepackage{float}
\usepackage{multirow}
\usepackage{color}
\usepackage{comment}
\usepackage{lipsum}
\usepackage{algorithm}
\usepackage{hyperref}
\usepackage[noend]{algpseudocode}
\usepackage{tikz}
\usepackage{verbatim}
\usepackage{subcaption}
\usepackage{enumitem}
\usepackage{paralist}
\usepackage[absolute]{textpos}

\begin{document}
\begin{textblock}{20}(1,1)
This is a preprint version. The paper is accepted by the 14th ACM International Systems and Storage Conference (Systor'21). \\
DOI: https://doi.org/10.1145/3456727.3463771
\end{textblock}

\title{IMG-DNA: Approximate DNA Storage for Images}
\author{
	Bingzhe Li \\
	School of Electrical and Computer Engineering\\
	Oklahoma State University\\
	Stillwater, OK, 74074\\
	\texttt{bingzhe.li@okstate.edu} \\
	\And
	 Li Ou \\
	Department of Pediatrics\\
	University of Minnesota, Twin Cities\\
	Minneapolis, MN, 55455 \\
	\texttt{ouxxx045@umn.edu} \\
		\And
	David Du \\
	Department of Computer Science and Engineering\\
	University of Minnesota, Twin Cities\\
	Minneapolis, MN, 55455 \\
	\texttt{du@umn.edu} \\
}

\maketitle	
\begin{abstract}
Deoxyribonucleic Acid (DNA) as a storage medium with high density and long-term preservation properties can satisfy the requirement of archival storage for rapidly increased digital volume. The read and write processes of DNA storage are error-prone. Images widely used in social media have the properties of fault tolerance which are well fitted to the DNA storage. However, prior work simply investigated the feasibility of DNA storage storing different types of data and simply store images in DNA storage, which did not fully investigate the fault-tolerant potential of images in the DNA storage. In this paper, we proposed a new image-based DNA system called IMG-DNA, which can efficiently store images in DNA storage with improved DNA storage robustness. First, a new DNA architecture is proposed to fit JPEG-based images and improve the image's robustness in DNA storage. Moreover, barriers inserted in DNA sequences efficiently prevent error propagation in images of DNA storage. The experimental results indicate that the proposed IMG-DNA achieves much higher fault-tolerant than prior work.

\end{abstract}

\section{Introduction}
The rapidly increased pace of digital volume motivates researchers to search for a storage medium with high areal density and long-term preservation. The International Data Corporation (IDC)~\cite{reinsel2018digitization} predicts the total amount of the whole world's digital data will reach 175 Zettabyte (ZB) in 2025. In addition to that, image data as one important data type is explosively increased since social media such as Facebook, Snapchat and Instagram have become more and more popular. Moreover, multiple resolution versions of images to support different requirements of devices and contexts further increase the demand for large scale storage systems~\cite{yan2017customizing}. Therefore, it is critical for emerging storage systems to store such a massive amount of image data with a low cost.

In past years, researchers have exploited many new emerging storage devices, such as Solid-State Drive (SSD)~\cite{lee2002effects}, shingled magnetic recording~\cite{amer2011data} and interlaced magnetic recording drives~\cite{wu2018data}, to satisfy the requirement of storage demand. However, these storage devices still face two challenges. One is that the increased storage density cannot catch up with the rapid rate of increased digital volume. The other is that the digital data can be reliably stored in these storage devices only for several years to at most tens of years. In other words, the data must be migrated to new drives for those years and the migration results in a much higher cost. Synthetic Deoxyribonucleic Acid (DNA) storage is one promising storage medium for archiving due to its high storage density and long durability. According to~\cite{appuswamy2019oligoarchive}, a theoretical density of DNA storage is about 455 Exabytes/gram and the data can be reliably stored in DNA for several centuries~\cite{allentoft2012half}. 

One issue of DNA storage is its high error rate. The write (synthesis) and read (sequencing) processes are error-prone. For each base pair (one nucleotide), it may involve around 1\% error rate~\cite{organick2018random}. To handle these errors, researchers use error-correction code (ECC) to recover errors resulting in a much high overhead. The space overhead may reach 15\% or even higher. According to the high error rate in DNA storage, images might be well fitted to the DNA storage due to its properties of fault tolerance and large volume in social media. Thus, DNA storage provides great potential to store a massive amount of image data to avoid high overhead induced by errors. 

There are two types of prior work related to how to store image data in DNA storage. One type of studies investigated the feasibility of DNA storage~\cite{church2012next}\cite{organick2018random}\cite{lopez2019dna}. Images as one type of applications are stored in DNA. They proposed different encoding schemes (i.e., converting digital data into DNA sequences), error-correction codes, and biochemical technologies for generally improving the efficiency of sequencing (read data out from DNA) or decreasing error rates. However, they did not focus on the property of image data, which cannot achieve the best performance for image data in terms of storage density and error tolerance. The other focused on approximately store images based on their property in different storage devices such as SSDs or Non-volatile memory~\cite{yan2017customizing}\cite{kuo2019long}\cite{fan2019adaptive}. For examples, Kuo~\textit{et al.}~\cite{kuo2019long} proposed a new JPEG encoding scheme to improve the error tolerance of SSD based on the sensitivities of JPEG image data. Fan~\textit{et al.}~\cite{fan2019adaptive} proposed an adaptive-length image coding scheme based on the reliability of NAND flash memory to improve the density and reduce the total cost of JPEG-based storage. However, due to the special properties of DNA storage (such as error propagation discussed in Section~\ref{sec:motivation}), those technologies cannot be efficiently applied in image-based DNA storage.  

In this paper, we proposed a new scheme to efficiently store JPEG-based images in DNA storage, called IMG-DNA, which combining the special properties of DNA storage and images. First, we propose a new DNA architecture by adding barriers to prevent error propagation in DNA storage. The barrier design is based on the encoding scheme and biochemical constraints in DNA storage, and thus it can be feasibly applied to DNA storage. Then, according to the error sensitivities of coefficients in JPEG images, we separately store those coefficients in different DNA sequences, and proper internal index is designed to associate these coefficients with their corresponding images. Moreover, to further improve the error tolerance of DNA storage, separate barrier schemes are applied to the different coefficients of JPEG-images according to their characteristics. Finally, the IMG-DNA scheme improves the error tolerance capability compared to prior work.

The remainder of this paper is organized as follows: Section~\ref{sec:background} describes the background of DNA storage and JPEG-based images. Section~\ref{sec:motivation} demonstrates the motivation behind this work. Section~\ref{sec:design} introduces the implementation of IMG-DNA scheme. Section~\ref{sec:results} shows the experimental results of IMG-DNA and compare them with some prior work. Related work is provided in Section~\ref{sec:related}. Finally, the conclusions are drawn in Section~\ref{sec:conclusion}.

\section{Background}\label{sec:background}
In this section, we mainly introduce the background of DNA storage and JPEG-based image.
\subsection{DNA Storage}
\begin{figure}[!t]
	\centering
	\includegraphics[width=3.3in]{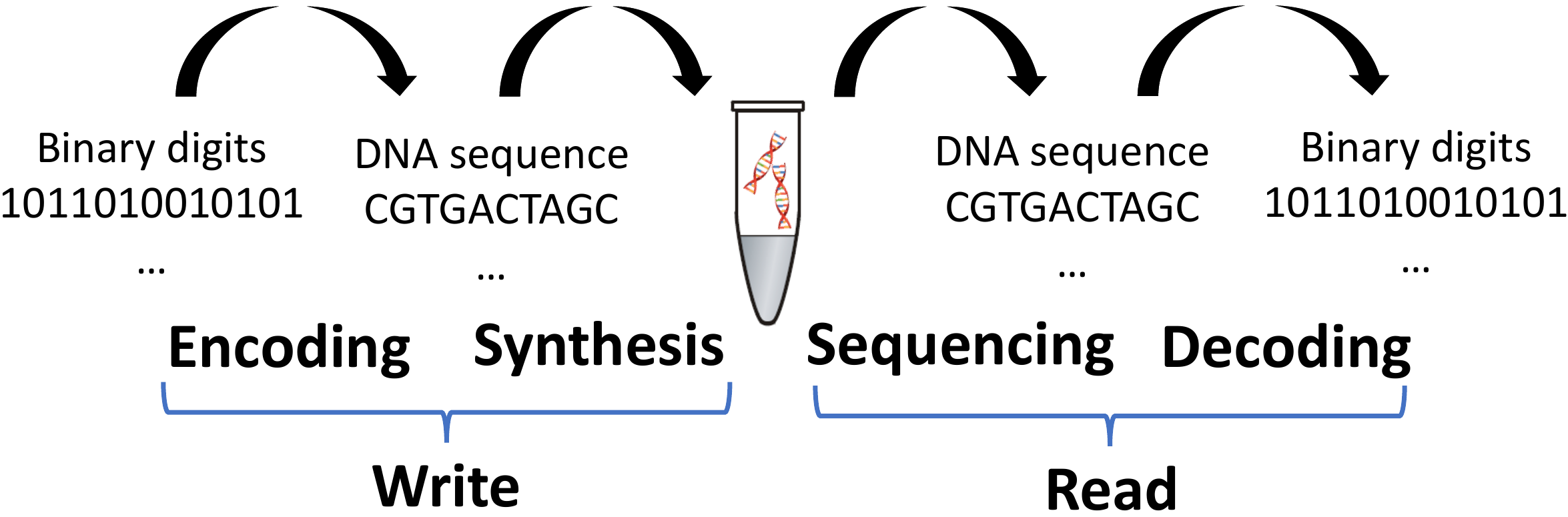}
	\caption{Basic steps of DNA storage.}
	\label{fig:basic_step}
\end{figure}

In DNA, four basic nucleotides (i.e., Adenine (A), Cytosine (C), Guanine (G), and Thymine (T)) consist of DNA sequences. Essentially, a DNA strand or oligonucleotide is composed of a number of nucleotides, and two DNA strands form a double helix. In this helix form, A and T are banded with each other, and C and G are aligned with each other. Therefore, these two DNA strands are complementary to each other. For DNA storage, as shown in Figure~\ref{fig:basic_step}, there are four major processes in the DNA storage system: encoding, synthesis, sequencing, and decoding. The encoding and synthesis are the processes to write digital data into DNA storage. The sequencing and decoding are the processes to read data out from DNA storage.

{\bf Encoding and decoding:} to enable storing binary digits to DNA storage, first of all, we should convert digital data into DNA sequence format (i.e., A, T, G, and C). Since there are four types of nucleotides in DNA, ideally each nucleotide can represent two binary bits (e.g., 00$\rightarrow$A, 01$\rightarrow$T, 10$\rightarrow$C and 11$\rightarrow$G). However, this two-bit encoding scheme introduces high errors in synthesis and sequencing processes due to biochemical constraints such as long homopolymers. In other words, some specific DNA patterns cannot be correctly read out, and we should avoid them in DNA storage such as long homopolymers (e.g., AAAAA). Therefore, most existing studies~\cite{organick2018random}\cite{church2012next}\cite{goldman2013towards}\cite{bornholt2016dna}\cite{erlich2017dna}\cite{blawat2016forward} achieve the encoding density less than 2 bits/nt (bit per nucleotide). To read data out from DNA storage, DNA sequences are decoded back to binary data according to encoding schemes.

{\bf DNA synthesis (write) and sequencing (read):} DNA synthesis and sequencing are two important processes to write and read data into and out from DNA sequences, respectively. For writing data in DNA storage, after encoding binary data to DNA sequences, we can chemically synthesize a DNA sequence nucleotide by nucleotide as designed in~\cite{kosuri2014large}\cite{matteucci1981synthesis}. After synthesis, millions of different DNA strands containing binary data are mixed in one tube/pool. During the read operation, the target DNA sequences are amplified/duplicated via Polymerase Chain Reaction (PCR). PCR is biotechnology for exponentially duplicating target DNA sequences within a tube. After that, the sample of DNA sequences is sent to a sequencing machine. The target DNA strands as a template will be duplicated by attaching fluorescent nucleotides with different colors. Finally, the target DNA sequence will be read out by the sequencing machine. These processes are error-prone since a nucleotide can be aligned to an existing partial strand. Current technologies can synthesize a DNA sequence up to 3,000 base pair (bp)~\cite{yazdi2015rewritable}\cite{GeneArt}. However, when the length is increased, the error rate happening on synthesis and sequencing processes also exponentially increased~\cite{erlich2017dna}\cite{yazdi2015rewritable}\cite{GeneArt}\cite{li2020can}. Due to these reasons, the majority of the existing works for DNA storage use  100\textasciitilde{}300 bp length of a DNA strand.

\begin{figure}[!t]
	\centering
	\includegraphics[width=3.3in]{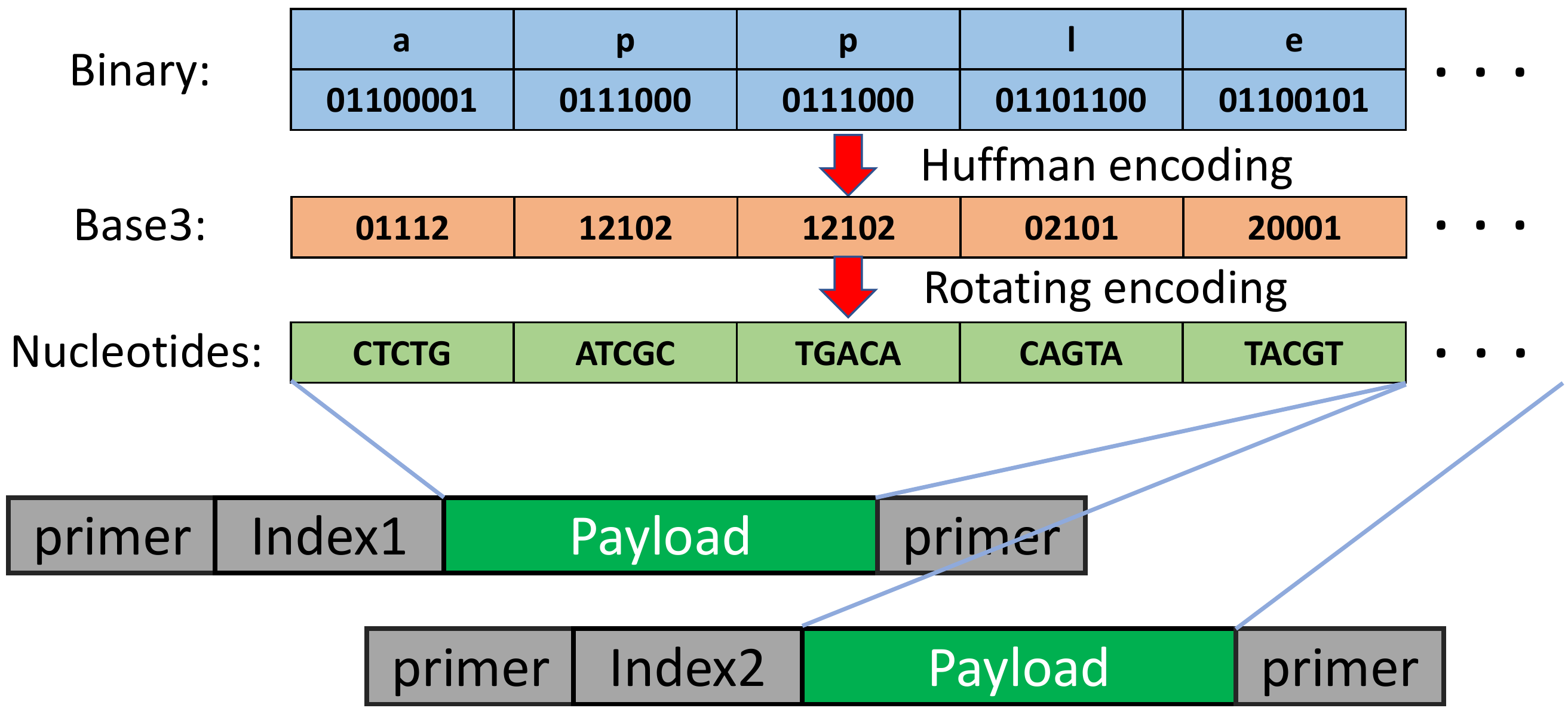}
	\caption{DNA format within the same primer pair.}
	\label{fig:DNA_process}
\end{figure}

\begin{figure}[!t]
	\centering
	\includegraphics[width=2.7in]{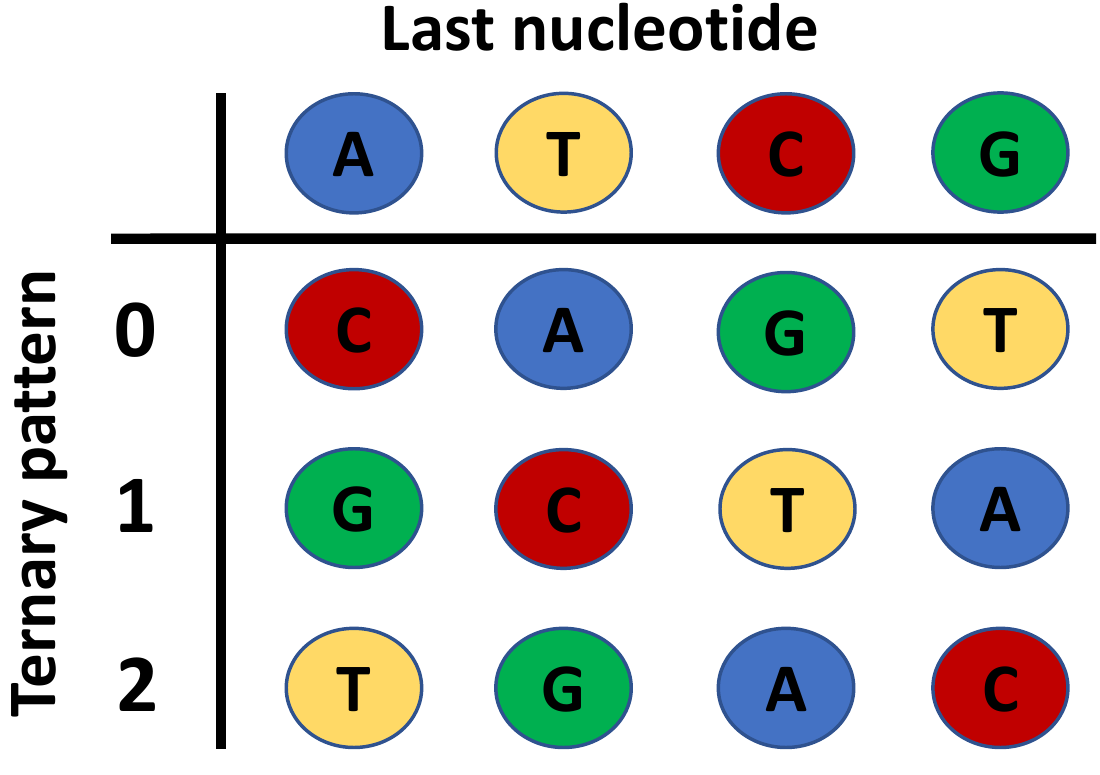}
	\caption{A rotating encoding to nucleotides avoids homopolymers (repetitions of the same nucleotide), which are error-prone.}
	\label{fig:DNA_encoding}
\end{figure}

{\bf Overall DNA storage processing: }
In DNA storage, one example of the basic DNA storage process~\cite{bornholt2016dna} is shown in Figure~\ref{fig:DNA_process}. First, a binary sequence is encoded with the base-3 Huffman code. For example, one eight-bit value can be encoded with 5 bits or 6 bits trits. The reason of using base-3 Huffman code is that the base-3 encoding manner can reduce the possibility of some errors in DNA storage compared to the base-4 encoding manner. Then, the base-3 sequence based on a rotating code in Figure~\ref{fig:DNA_encoding} is converted to a DNA sequence. The rotating manner can avoid long homopolymers (e.g., AAAAA) in DNA storage, which are error-prone. After that, the DNA sequence will be chunked into fixed-size payloads. The payload associated with primers and internal index consists of a DNA strand. A primer is a short nucleotide attached to the starting or ending points of DNA strands for DNA sequencing and PCR ranging from 18\textasciitilde{}25 bp~\cite{organick2018random}. One primer pair (i.e., we need two different primers for a DNA strand) can be associated with thousands or millions of different DNA sequences. Therefore, to distinguish these DNA sequences, an internal index field is used to distinguish all these DNA sequences. ECC (error-correction code) is typically used for recovering original data and might encode multiple DNA sequences to generate a new one~\cite{bornholt2016dna}.

\subsection{Fundamentals of JPEG-based Image}
JPEG (Joint Photographic Experts Group), as one of the most important image formats, is widely used in many fields such as social media, digital cameras, and smart phones. It occupies a high percentage of websites using various image file formats~\cite{JPEG_market}. In this paper, we mainly focus on the JPEG-based encoding for images.

During encoding, first an image is divided into blocks with 8*8 pixels. After that, the JPEG encoding scheme employs the discrete cosine transform (DCT) to compress each block from the spatial domain to the frequency domain. Through the DCT process, pixels at the left-top side are low-frequency components, and the pixels at the right bottoms are high-frequency components. Then, to further compress the image, since the human eyes can easier recognize the low-frequency components compared to the high-frequency spatial region, a quantization process is applied to assign different resolutions to the low-frequency component (DC) and the high-frequency components (AC). These components are divided by the non-uniform entries of a quantization matrix. The DC part achieves a higher resolution than AC components.

\begin{figure}[!t]
	\centering
	\includegraphics[width=3.3in]{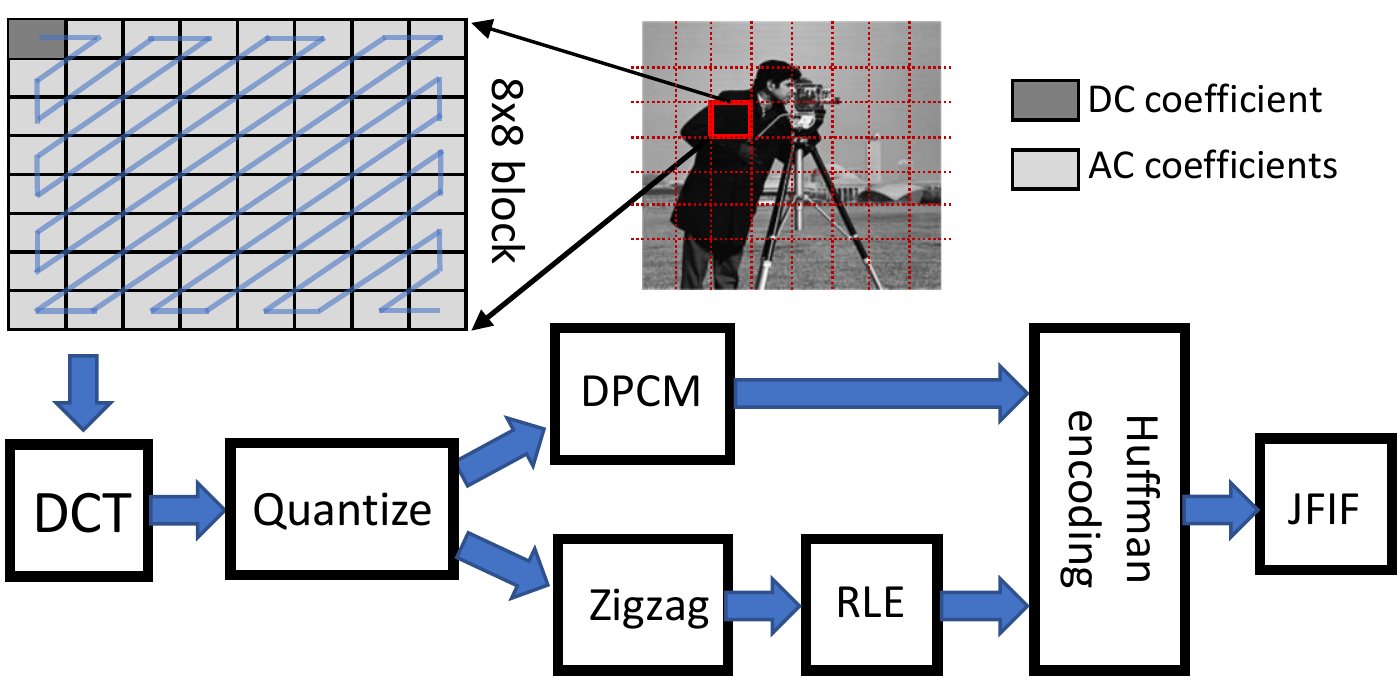}
	\caption{The encoding process of JPEG-based image.}
	\label{fig:jpeg}
\end{figure}

Following that, there are two primary paths for the JPEG encoding scheme as shown in Figure~\ref{fig:jpeg}. For the DC path, the DC coefficient of each block goes through differential pulse code modulation coding (DPCM) to compute the difference between two consecutive DC components. For the AC path, first AC coefficients are transferred to a sequence value based on a zigzag order. Then, the AC zigzag sequence is further compressed by a run-length encoding (RLE)~\cite{yang2008joint}. Finally, different Huffman codes are applied to DC and AC coefficients separately~\cite{ghanbari2003standard}, and all those coefficients are encapsulated to the JPEG File Interchange Format standard (JFIF).

\section{Motivation}\label{sec:motivation}
In this section, we introduce the motivation for this work. Images as one of the most important data types are widely stored and used in cloud storage and social media. There are two properties of image-based data. One is the large volume since image-based data type is explosively increased and generated from social media such as Facebook, Snapchat, and Instagram. The large volume increases the demand for high-capacity storage systems. The other property of image-based data is fault tolerance. Many existing studies investigated images in approximate storage such as flash memory or non-volatile memory to reduce the storage overhead or improve the robustness of images~\cite{kuo2019long}\cite{fan2019adaptive}\cite{li2019leveraging}. Therefore, these two properties of images are well fitted to the DNA storage system, which is error-prone and has an extremely high density. However, none of the previous studies focused on the robustness of images in DNA storage. Although some existing studies~\cite{organick2018random}\cite{church2012next}\cite{bornholt2016dna} implemented image-based binary data in DNA storage, all of them focused on the feasibility of DNA storage for different types of data including images. Therefore, how to efficiently store images and improve the robustness of images in DNA storage is a critical and interesting research issue.

\begin{figure}[!t]
	\centering
	\includegraphics[width=3.3in]{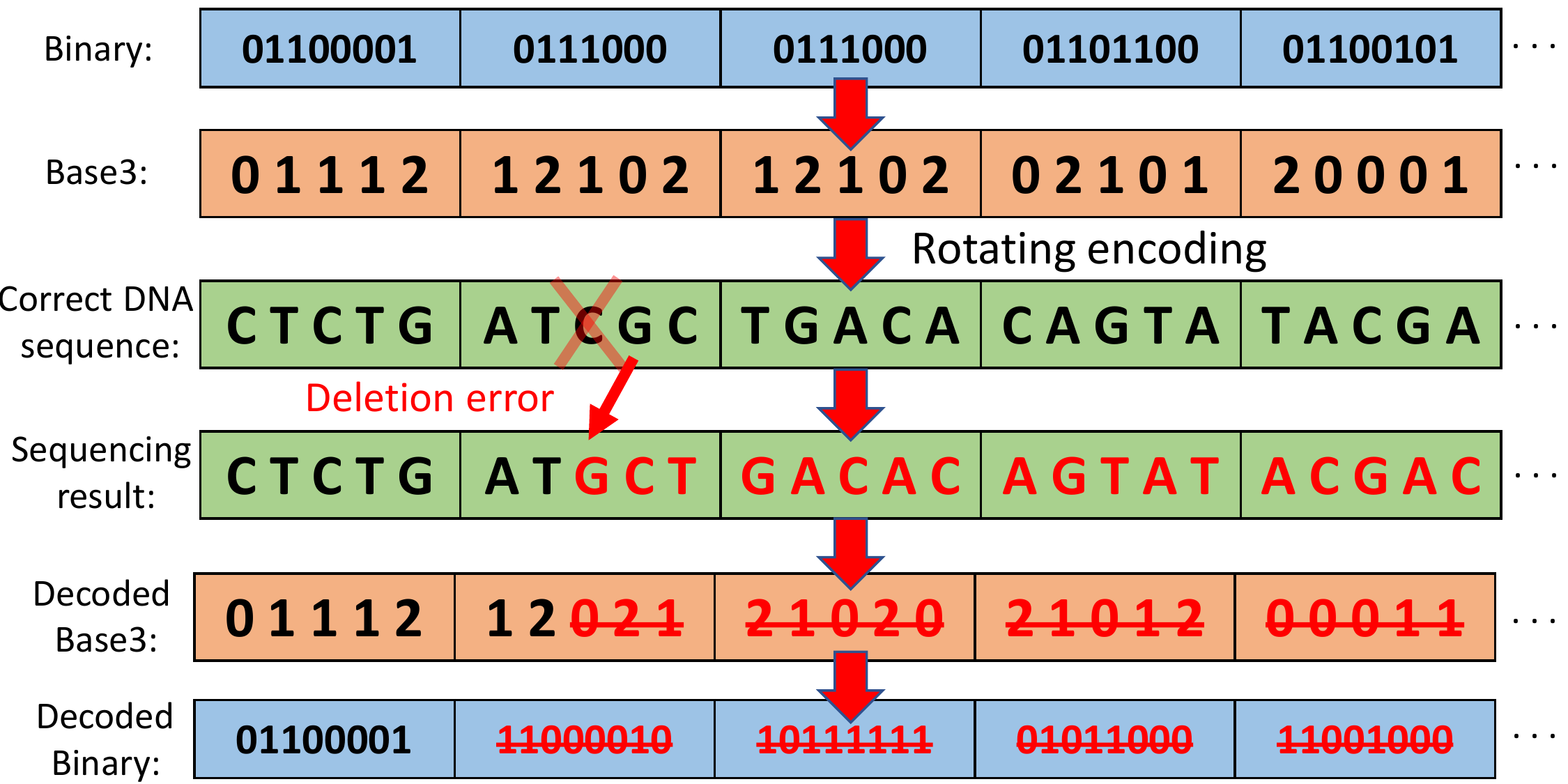}
	\caption{An example of deletion error for error propagation. The deletion error causes that one nucleotide `C' is missed and all subsequence are decoded wrongly.}
	\label{fig:error_pro}
\end{figure}

Moreover, due to the biochemical constraints as mentioned before, we should avoid long homopolymers in DNA sequences, which cause high error rates. Therefore, the state-of-the-art encoding scheme~\cite{church2012next}\cite{goldman2013towards}\cite{organick2018random}\cite{blawat2016forward} uses a rotating code manner as shown in Figure~\ref{fig:DNA_encoding} to avoid long homopolymers. That is, the current nucleotide is based on the current digit and its last encoded nucleotide. For example, as shown in Figure~\ref{fig:DNA_process}, the first nucleotide of the input `e' (i.e., `01000101' in ASCII and `20001' in base-3 Huffman code) is encoded to 'T' based on the current digit '2' and its last nucleotide 'A'. However, the rotating manner induces a phenomenon called error propagation (DNA-level). In DNA storage, there are three types of errors (i.e., insertion, substitution, and deletion). Any nucleotide changed in the middle of one DNA strand has a significant influence on its following nucleotides. As a result, all subsequence data cannot be correctly read out due to one nucleotide error. As shown in Figure~\ref{fig:error_pro}, a deletion error occurs in the middle of one DNA sequence and causes a series of errors for its following nucleotides. Such error propagation significantly affects the robustness of the image-based storage system. Additionally, although prior work~\cite{kuo2019long}\cite{fan2019adaptive} has mitigated the image-level error propagation for traditional storage devices, these techniques are not efficient for the DNA-level error propagation due to the special encoding manner in DNA storage. Therefore, there is a need to find a proper architecture preventing such error propagation in DNA storage.

\section{Algorithm Design}\label{sec:design}
In this section, we introduce the proposed IMG-DNA scheme to improve the robustness of image-based DNA storage. The IMG-DNA scheme has three major optimizations compared to previous work. The first one is that the separation of DC and AC coefficients at the DNA level. The second one is `adding barrier' to both AC and DC coefficients. The third one is that the DC and AC coefficients are applied with different densities of barriers. In the following subsections, we introduce each optimization in detail.

\subsection{AC/DC Coefficient Separation at DNA Level}
According to JPEG-based images, the DC and AC coefficients have different sensitivities to errors. There are two reasons. One is that the DC coefficients have higher resolution than the AC coefficients during the quantization process since the human visual system is more sensitive to DC coefficients (low-frequency part). Therefore, under the same error rates, the DC coefficients have more significant influences on the quality of an image. The other reason is that the DC coefficients use differential pulse code modulation coding (DPCM) to compute its consecutive DC value for the next 8*8 block. So, if the DC coefficient of the first block is injected an error, the coefficients in the following blocks are decoded based on a wrong value, resulting in much high errors, which is called error propagation phenomenon~\cite{kuo2019long}. Due to the different sensitivities of DC/AC coefficients, it is better to separately store them to avoid influences with each other. The DNA storage system provides the possibility of physical separation. We can store DC/AC coefficients in different DNA strands.  To distinguish them, the internal index can separate these coefficients during sequencing and decoding processes. The details can be found in Section~\ref{sec:overall_DNA}. Moreover, the DC/AC separation provides a possibility of using different robustness schemes for DC/AC coefficients as discussed in Section~\ref{sec:DC/AC_difference}.

\subsection{Adding `Barriers'}
In this subsection, we introduce the scheme of adding `Barriers' to prevent the error propagation phenomenon as discussed in Section~\ref{sec:motivation}. The error propagation is caused due to the errors induced by the synthesis and sequencing processes as shown in Figure~\ref{fig:error_pro}. We can use some 'barriers' to prevent such error propagation. However, the barriers are not easy to be added since the barriers should be distinguished from the payload.

\begin{figure}[!t]
	\centering
	\includegraphics[width=3.3in]{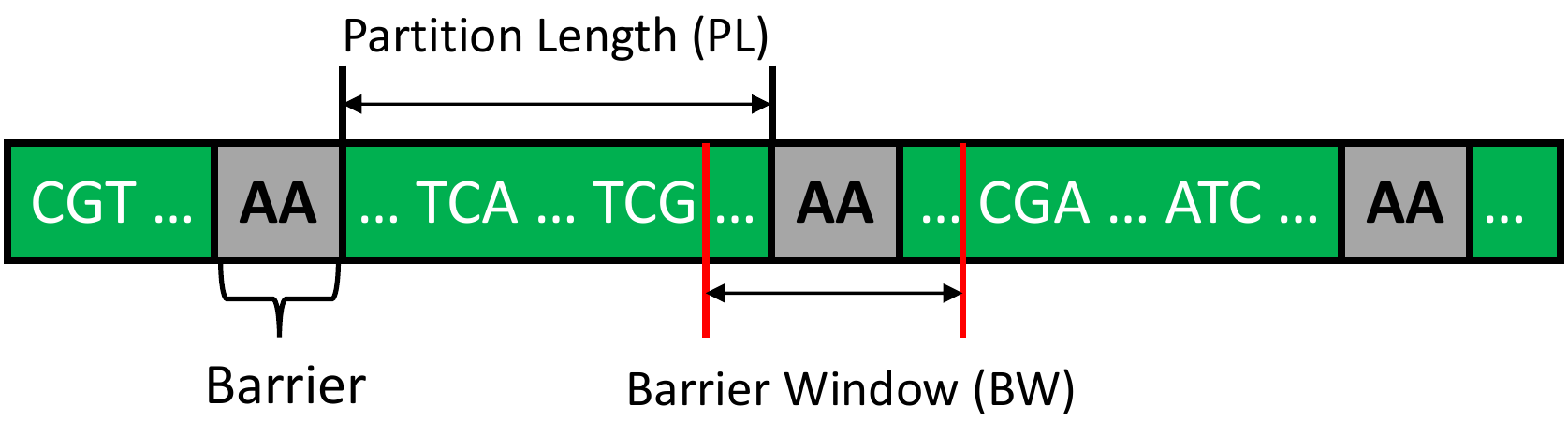}
	\caption{Definition of Partition Length (PL) and Barrier Window (BW). }
	\label{fig:barrier}
\end{figure}
The intuition of adding `Barriers' to prevent error propagation is to break the relationship between nucleotides. Moreover, the `Barriers' should be easily distinguishable with other nucleotides. According to the encoding scheme, the rotating encoding scheme can avoid any two consecutive identical nucleotides as shown in Figure~\ref{fig:DNA_encoding}. As a result, the biochemical constraints such as long homopolymers can be avoided. The constraint of the number of consecutive identical nucleotides is less than four nucleotides~\cite{blawat2016forward}. Therefore, since there are no two consecutive identical nucleotides based on the rotating coding scheme and two consecutive identical nucleotides are allowable, we use the pattern 'AA' as a barrier to partition the DNA sequence with a fixed sequence length (e.g., 50 bp) named Partition Length (PL) as shown in Figure~\ref{fig:barrier}. So, the digital values are partitioned with a length of N values (i.e., PL = N), and then the digital values are converted into the DNA sequence individually. After that, we insert `AA' barriers in between N nucleotides. It is possible that three consecutive `A's are generated in the final DNA sequence if the inserted position has an `A' just before the barrier since we use the encoding as the rotating manner. The reason of using the barrier pattern based on `A' is that the `AA' and `AAA' patterns have higher sequencing accuracies than other three types of patterns~\cite{ivady2018analytical}. Since we predefine the PL, an error will be prevented within its partition and will not affect other partitions. So, one error causes the maximum errors within a partition size (i.e., PL).

\begin{figure}[!t]
	\centering
	\includegraphics[width=3.3in]{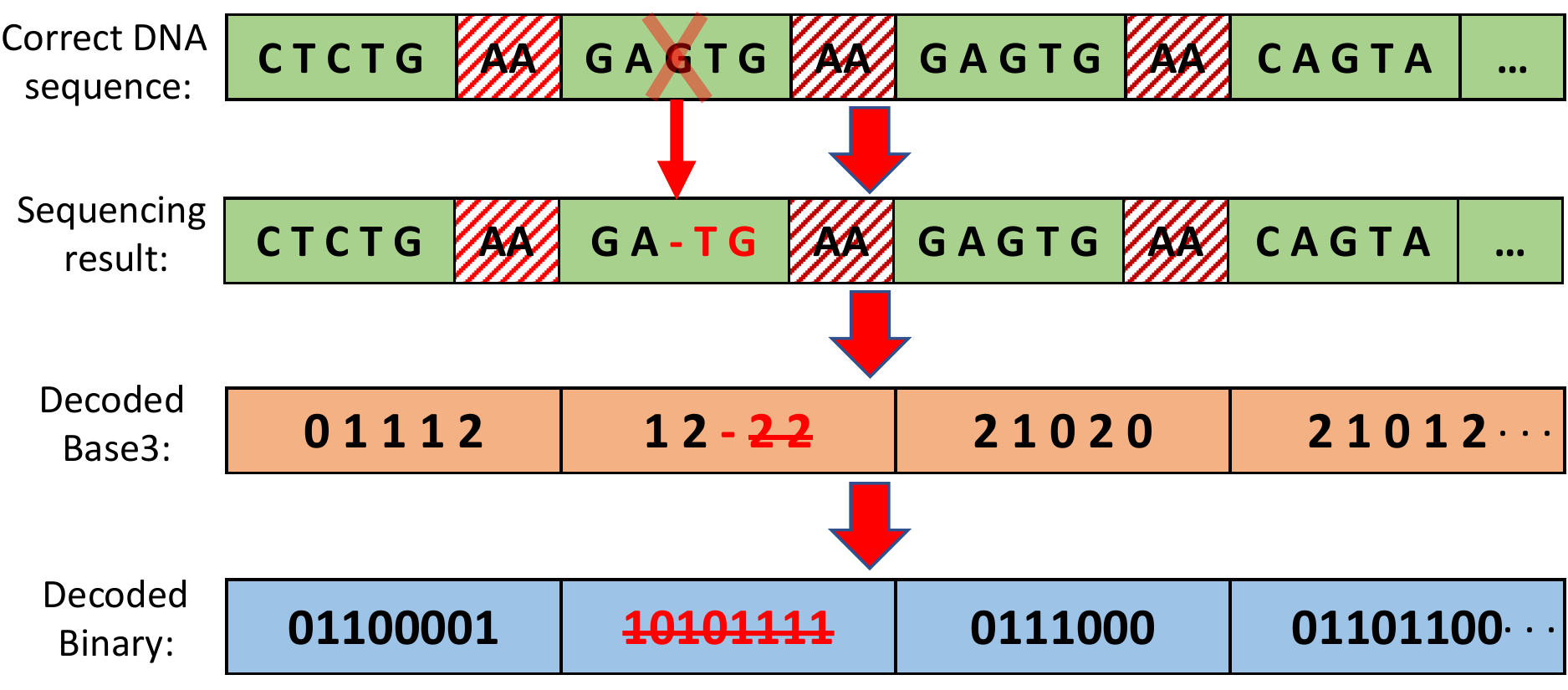}
	\caption{An example of adding `Barriers' to prevent the error propagation from a deletion error.}
	\label{fig:error_pro2}
\end{figure}

Figure~\ref{fig:error_pro2} shows an example of adding `Barriers' to prevent the error propagation from a deletion error. Compared to Figure~\ref{fig:error_pro}, the proposed IMG-DNA adds `AA' to break the relationship between different nucleotides. Also, we know that the number of nucleotides between two barriers `AA' is a constant number (i.e., PL = 5 in this example). So, a deletion error in the second partition can be detected. Finally, the deletion error only happens within the partition. For other types of errors (e.g., substitution and insertion), adding `Barriers' can achieve similar effectiveness of the prevention of error propagation like the example in Figure~\ref{fig:error_pro2}.

Moreover, substitution, deletion, and insertion errors may generate a pattern which is the same as the barrier pattern (i.e., AA). Misclassification of the barrier patterns may mitigate the effect of error propagation prevention. To mitigate such a scenario, we propose another mechanism called Barrier Window (BW), which means that only a pattern 'AA' locating in a barrier window will be regarded as a true barrier if detected. The pattern 'AA' outside the barrier window will be treated as a normal payload. If there is no barrier pattern in the BW, it means that an error happens on the barrier pattern and changes the barrier to other patterns. Thus, we increase the number of partitions by one and use the next barrier pattern to prevent the error propagation. In this case, the error will be propagated to the next partition but we can prevent it in two partitions. We set BW to 12 bp, which means 5 bp on both left and right sides of the barrier pattern. Since the probability of errors happening on the barrier pattern is much lower than that of other payloads (25 times less if the fixed-length takes 50 bp), it is highly possible that the error propagation can be prevented in one or two partitions. Therefore, with adding barriers and Barrier Window, the IMG-DNA scheme can significantly prevent the error propagation in image-based DNA storage.  

\subsection{Asymmetric Barriers for AC/DC Coefficients}\label{sec:DC/AC_difference}
As discussed above, the barriers efficiently prevent error propagation within one partition. Thus, the image-based DNA storage with a shorter partition length (PL) can achieve higher robustness but it introduces a larger capacity overhead. Too many barriers are added DNA sequence resulting in a low encoding density. However, if we take a too large PL, the robustness of DNA storage is decreased. Therefore, there is a trade-off between the robustness and encoding density of DNA storage.

To balance the robustness and encoding density of DNA storage, we propose an asymmetric design for image-based DNA storage. According to the properties of JPEG-based images as discussed in Section~\ref{sec:DC/AC_difference}, the DC coefficients are more sensitive to errors than AC coefficients and the total size of DC coefficients are much smaller than that of AC coefficients (about 63 times less). So, we use an asymmetric barrier design for AC/DC coefficients. Since DC coefficients are more sensitive to errors, we assign a smaller PL value to DC coefficients to increase the robustness of DNA storage. Meanwhile, since AC coefficients have much larger size than that of DC, a larger PL value for AC coefficients can reduce the overhead of the scheme of adding barriers. In this paper, we use 20 bp for DC PL and 50 bp for AC PL by default. By doing so, a shorter PL value in DC coefficients can improve the robustness of image-based DNA storage significantly. Moreover, compared with the data sizes of DC and AC coefficients in JPEG images, normally DC coefficients only occupy 4\% - 8\% of one image size~\cite{fan2019adaptive}. Thus, the increased overhead of adding more barriers in DC coefficients is limited. According to our experiments, the capacity overheads of barriers in DC and AC coefficients are about 0.28\% and 3\% of image size with PL=20 bp for DC and PL=50 bp for AC, respectively.

\subsection{Image-based DNA Storage Architecture}\label{sec:overall_DNA}
\begin{figure}[!t]
	\centering
	\includegraphics[width=3.3in]{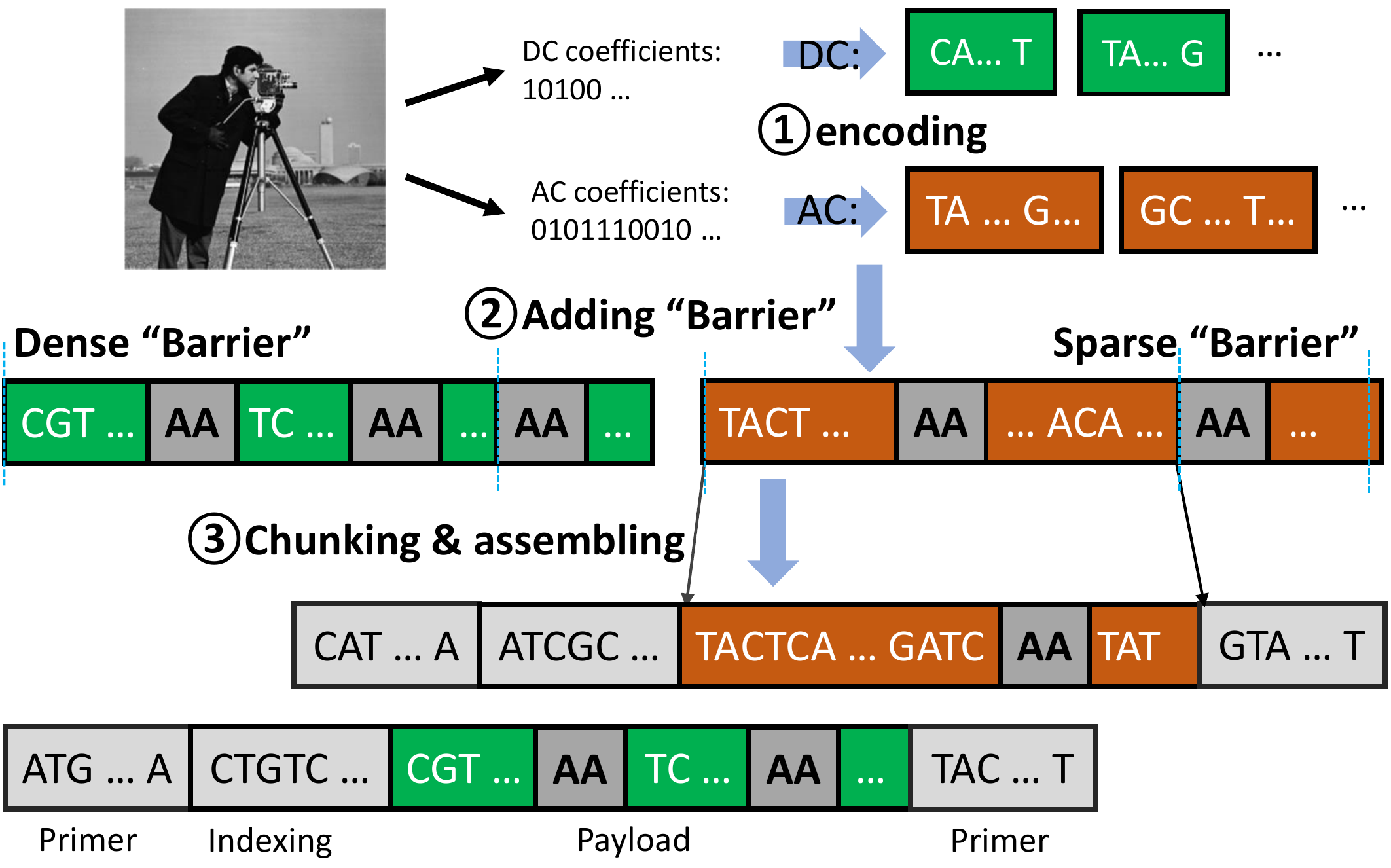}
	\caption{The overall architecture of proposed IMG-DNA.}
	\label{fig:DNA_architecture}
\end{figure}
Figure~\ref{fig:DNA_architecture} provides an overall architecture of proposed IMG-DNA. After JPEG-based encoding, we first separate DC and AC coefficients. This step provides the opportunities to apply the asymmetric barrier scheme in the DNA storage encoding process. After that, the digital sequence will be encoded into DNA sequence. Then, we insert the barrier pattern 'AA' into DC and AC coefficients separately. According to the different error sensitivities of DC and AC, we use different partition lengths for AC and DC DNA strands. After inserting barriers, long DNA sequences will be chunked into small segments with a fixed length (e.g., 200 bp). Then, each segment with a primer pair and its corresponding internal index are assembled. For the metadata of images including image height and width, precision, Huffman table, etc., a tiny error in those metadata will cause a significant effect on the quality of images. So, similar to~\cite{kuo2019long}, we store those metadata with a full protection by ECCs in either DNA storage or a traditional storage medium. In other words, there are no errors when reading metadata out. Since the size of metadata is much smaller than that of AC/DC coefficients, the overhead of correctly storing and retrieving metadata is acceptable.

Moreover, there is a two-level indexing in DNA storage systems~\cite{li2020can}. One is to index DNA strands in the same tubes. The other is a file/object level indexing to indicate which DNA strands belong to one specific image. The internal index as shown in Figure~\ref{fig:DNA_architecture} contains the type and address offset of payload. So, the internal index can distinguish the DNA strands associated with the same primer pair and also distinguish the AC/DC coefficients. For the file/object level indexing, a mapping table needs to be preserved to indicate the relationship between images and DNA strands such as an image ID pointing to its corresponding primers and its internal offsets.

\subsection{Feasibility Discussion}\label{sec:feasibility}
In this study, the experiments are based on simulation. Although the validation of this work does not do the wet-lab experiment, which will be left for future work, the software-based feasibility check is a common technique in biological fields such as primer design. The simulation result can have a high correlation with the wet-lab experiments and be able to reflect the success rate of wet-lab experiments. 

A set of design rules for synthesis and sequencing efficiency are used based on commercial design rules~\cite{DNA_rules, Synthesis_rule} and previous studies~\cite{bornholt2016dna, erlich2017dna, organick2018random}. Three major design rules need to be followed. One is that the absence of long homopolymers (less than four nucleotides)~\cite{erlich2017dna, blawat2016forward}. The second one is that GC contents in a sequence should be around 40\% - 60\%. The third one is DNA strand length smaller than 1000 bp. For the proposed IMG-DNA, it satisfies all these three design rules. The IMG-DNA at most has three homopolymers due to adding barriers. Moreover, due to the rotation encoding manner and short DNA sequence length for DC and AC coefficients, the IMG-DNA can keep the GC contents (i.e., around 49\%) within 40\% - 60\% and make the DNA sequence length shorter than 1000 bp. Therefore, according to the above discussion, we can conclude that the IMG-DNA is feasible to be used in DNA storage systems for synthesis and sequencing as previous studies~\cite{bornholt2016dna, organick2018random}.

\begin{figure*}[!t]
	\centering
	\setlength{\lineskip}{\medskipamount}
	\subcaptionbox{Original}{\includegraphics[width=1.7in]{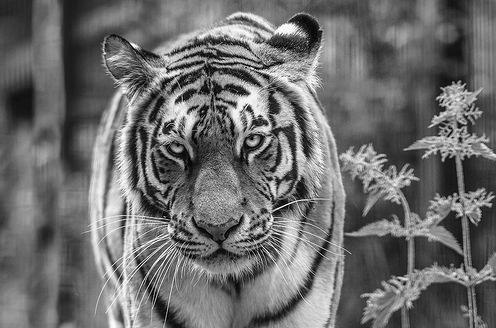}}\hfill
	\subcaptionbox{IMG-DNA (SSIM=0.9078)}{\includegraphics[width=1.7in]{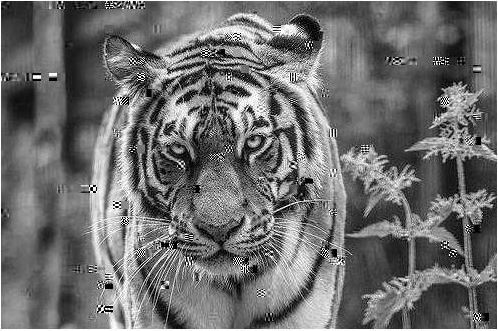}}\hfill
	\subcaptionbox{Approx-DNA (SSIM=0.1604)}{\includegraphics[width=1.7in]{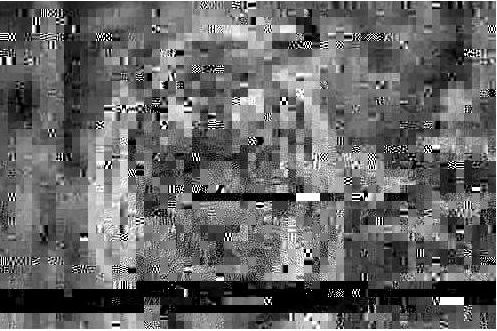}}\hfill
	\subcaptionbox{Raw-DNA (SSIM=0.0561)}{\includegraphics[width=1.7in]{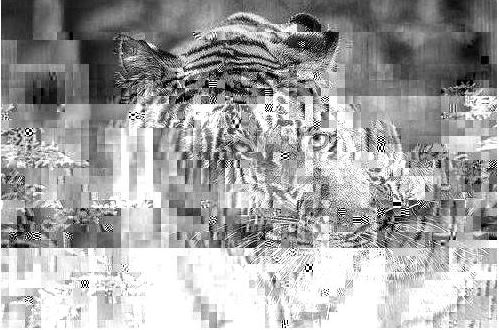}}\hfill
	\caption{Graphic view of different schemes with injecting 0.1\% error rate for all coefficients.} \label{fig:graph_error_injection}
\end{figure*}
\begin{figure}[!t]
	\centering
	\includegraphics[width=3.3in]{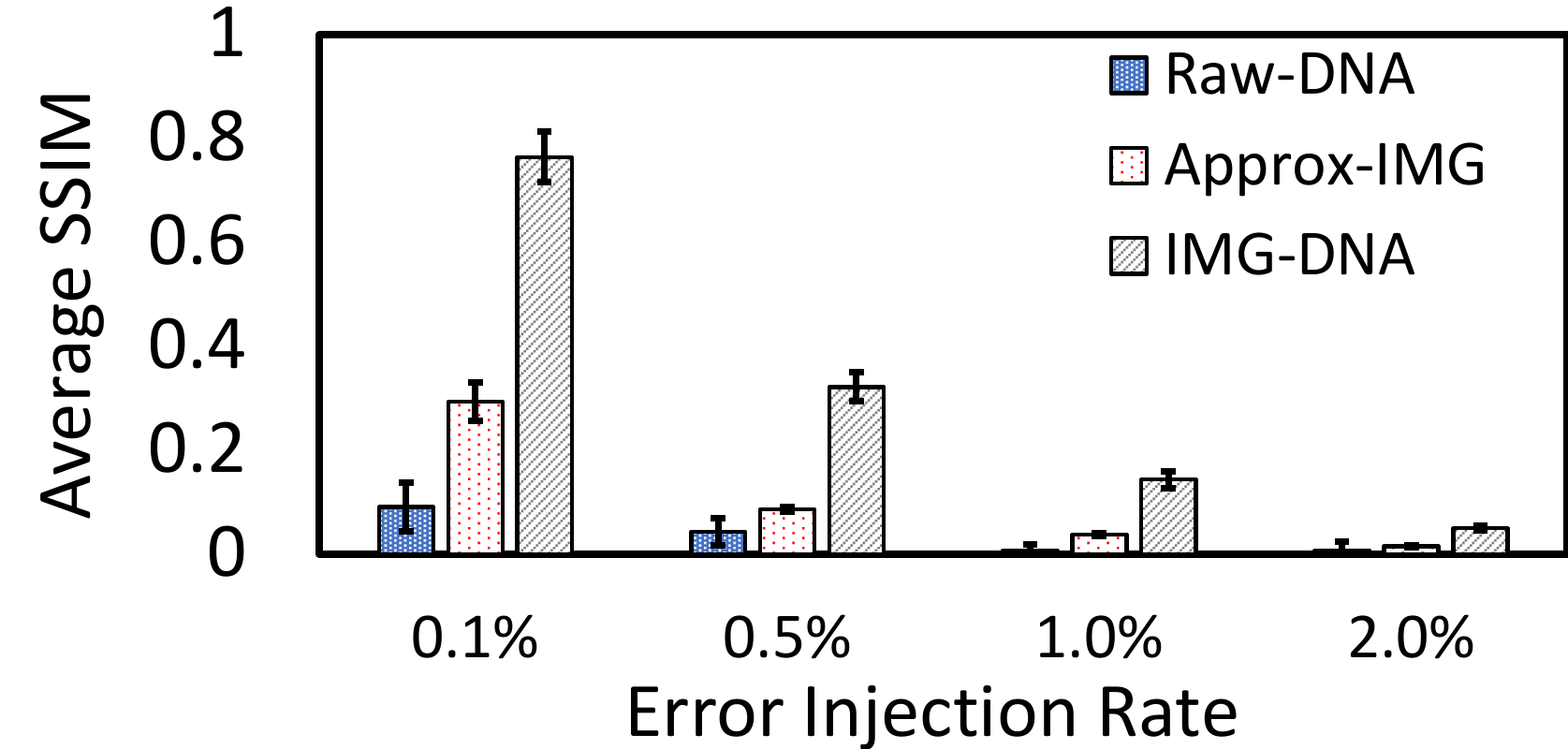}
	\caption{SSIM comparison with injecting errors into both AC and DC coefficients with a 90\% confidence interval.}
	\label{fig:overall_error}
\end{figure}
\section{Experimental Results}\label{sec:results}
In this section, we investigate the robustness of the DNA storage system for storing images. The default DNA strand length is 250 bp including a primer pair, payload, and index (the DNA strand lengths may be changed slightly due to the length of the index). Two previous studies are used as baselines. One is the straightforward implementation of DNA storage~\cite{organick2018random}\cite{bornholt2016dna} (denoted by \textbf{Raw-DNA}). The other one is the approximate image schemes with the digital-level optimization~\cite{kuo2019long} (denoted by~\textbf{Approx-IMG}). We use 500 JPEG-based images from the ImageNet image dataset~\cite{imagenet} for all experiments. The error model is extracted from error distributions of the wet-lab DNA storage implementation~\cite{organick2018random}. Our experiments use MATLAB2020a to encode real images into DNA sequences based on different schemes, with running in a system with Intel i-7-47900 CPU@3.6GHz and 8GB memory. To investigate the robustness of DNA storage, we use a metric called SSIM (structural similarity index metric)~\cite{wang2004image}, which is widely used to quantify the similarity between the image with no errors and the images with injected errors. The value of SSIM is between -1 to 1. The larger SSIM value indicates that two images are more similar to each other. So, if two images are near-identical, the SSIM will be close to 1.

\subsection{Robustness of Image-based DNA System}
We investigate the robustness of image-based DNA systems for the ImageNet data set by manually injecting errors from 0.1\% to 2\% following the distribution of the error model~\cite{organick2018random}. The AC and DC coefficients use the same DNA strand length (250 bp) and different partition lengths (50 bp and 20 bp respectively). As shown in Figure~\ref{fig:overall_error}, for all three schemes, the average SSIMs trend to be decreased as injecting more errors. Compared with these three schemes, the proposed IMG-DNA achieves a much higher fault tolerance ability than the other two with different error rates (2.6x - 19.7x SSIM improvement). The reason is that the IMG-DNA considers both digit- and DNA- level error propagation prevention for images. Additionally, a 90\% confidence interval is added to each bar in Figure~\ref{fig:overall_error} to indicate the SSIM variance among different images. Based on the results, we can find that even though SSIM might be changed for different images, the variance is not large for all images and the proposed IMG-DNA can always achieves higher robustness than others.

Moreover, a graphic view of an image with different encoding schemes are used in Figure~\ref{fig:graph_error_injection}. We can find that the proposed IMG-DNA achieves much close vision to the original picture and the other two baselines are vague on parts of the picture.

In summary, without any optimization, the propagation error in DNA storage can cause image corruption even with a small error in AC or DC data. By using the proposed scheme, the error propagation is mitigated and finally results in little image quality degradation. The proposed scheme can assist those protection schemes and reduce the overhead of those ECCs (error-correction codes). For example, originally one error can propagate to ten errors. Thus, to recover all those ten errors, a strong ECC should be used. With the prevention of the error propagation, the ECC only needs to recover this one error. Moreover, removing ECCs can significantly improve the capacity for approximate storage. Thus, in this paper, only metadata is protected by ECC and AC/DC coefficients are encoded to DNA sequences without using ECC.

\subsection{Effect of Different Partition Lengths}
\begin{figure}[!t]
	\centering
	\includegraphics[width=3.3in]{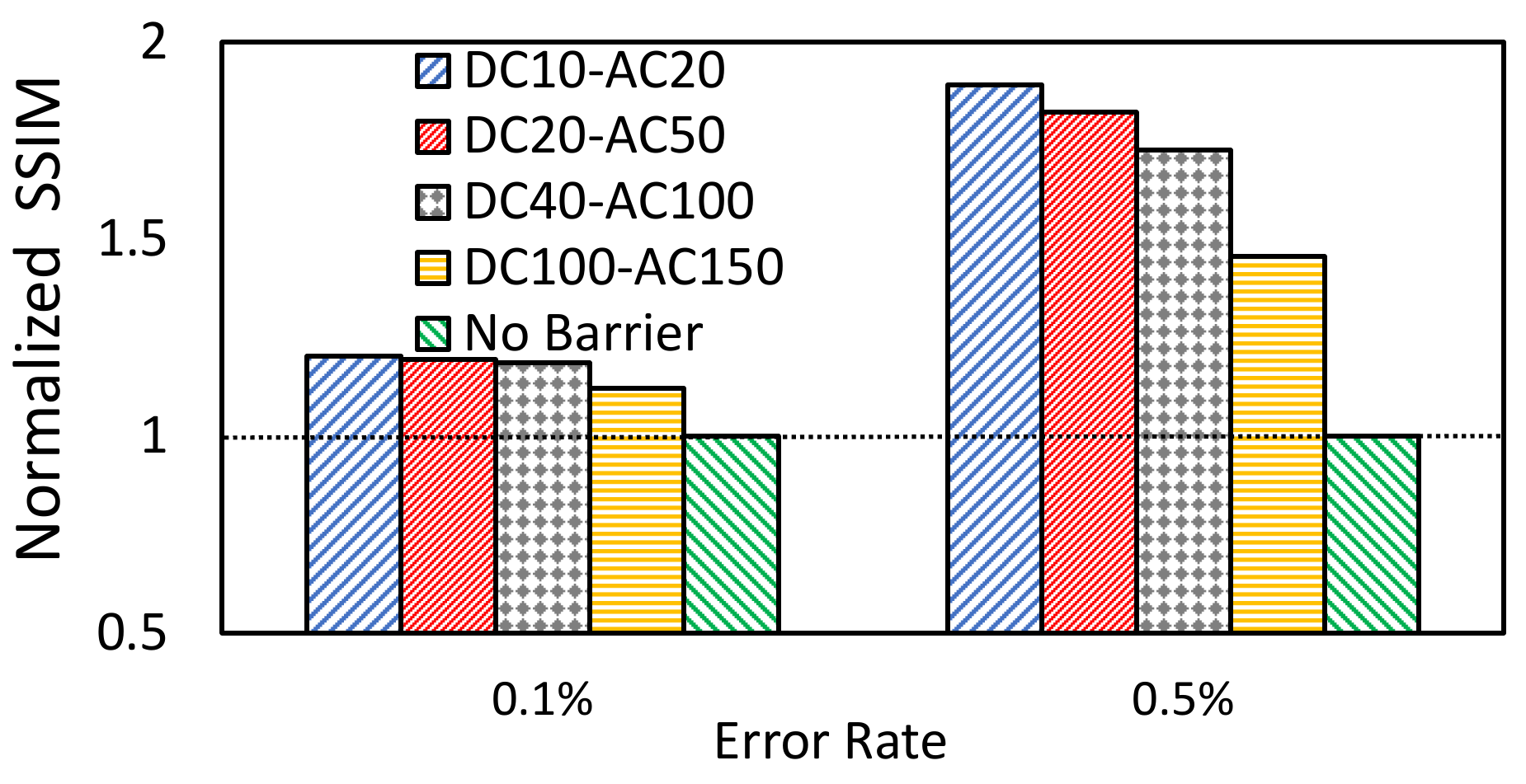}
	\caption{Normalized SSIM with different PL values. (DC20-AC50 refers to DC with PL=20 and AC with PL=50).}
	\label{fig:PL_varying}
\end{figure}
We compare SSIM values by varying the PL values for AC and DC coefficients. The IMG-DNA scheme with no barriers is normalized to 1 (denoted by `No Barrier'). As shown in Figure~\ref{fig:PL_varying}, as the PL values increase for AC and DC, the SSIM is decreased. In other words, the robustness of image-based DNA storage becomes lower with larger PL values. Moreover, when injecting error rates increase, the SSIM difference between PL values become larger. Therefore, for the cases with high error rates, it is always better to use a smaller PL value to prevent errors while may introduce a higher capacity overhead.

\subsection{Errors on Different Coefficients}
\begin{figure}[!t]
	\centering
	\includegraphics[width=3.3in]{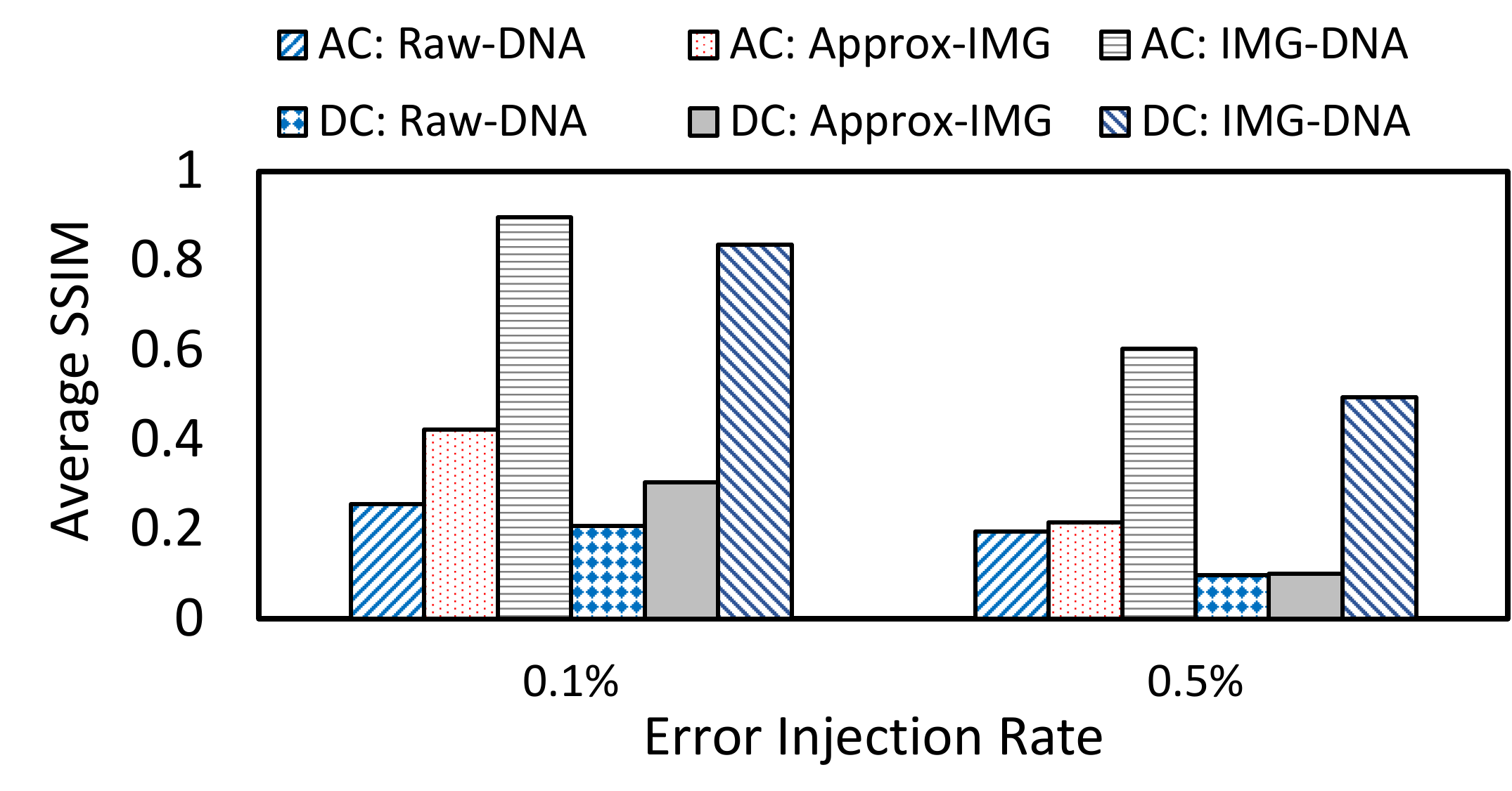}
	\caption{SSIM comparison with separately injecting errors into AC and DC coefficients.}
	\label{fig:error_DCAC}
\end{figure}

In this subsection, we investigate the effect of error injection on different coefficients. As indicated in Figure~\ref{fig:error_DCAC}, for all three schemes, as injecting different errors on AC and DC coefficients, injecting errors in DC coefficients has a much larger effect on the quality of images than that of AC coefficients. As mentioned in the background section, the reason is that the DC coefficients are the mean values of image blocks and thus any change of the DC coefficients directly affects the whole blocks. If we compare different schemes, we can find that the proposed IMG-DNA can support much higher SSIM values for both AC and DC coefficients than the other two schemes. The reason is that the proposed scheme of adding barriers prevents error propagation in DNA storage and thus improves the overall robustness of image-based DNA storage systems.

\subsection{Overhead Discussion}
In this subsection, we discuss the overhead of the IMG-DNA scheme. There are two major types of overhead, computation overhead and capacity overhead. For the computation overhead, since we use a similar rotation encoding scheme as~\cite{organick2018random}, the computation overhead including encoding and decoding latencies of both schemes is much similar. 

\begin{figure}[!t]
	\centering
	\includegraphics[width=3.3in]{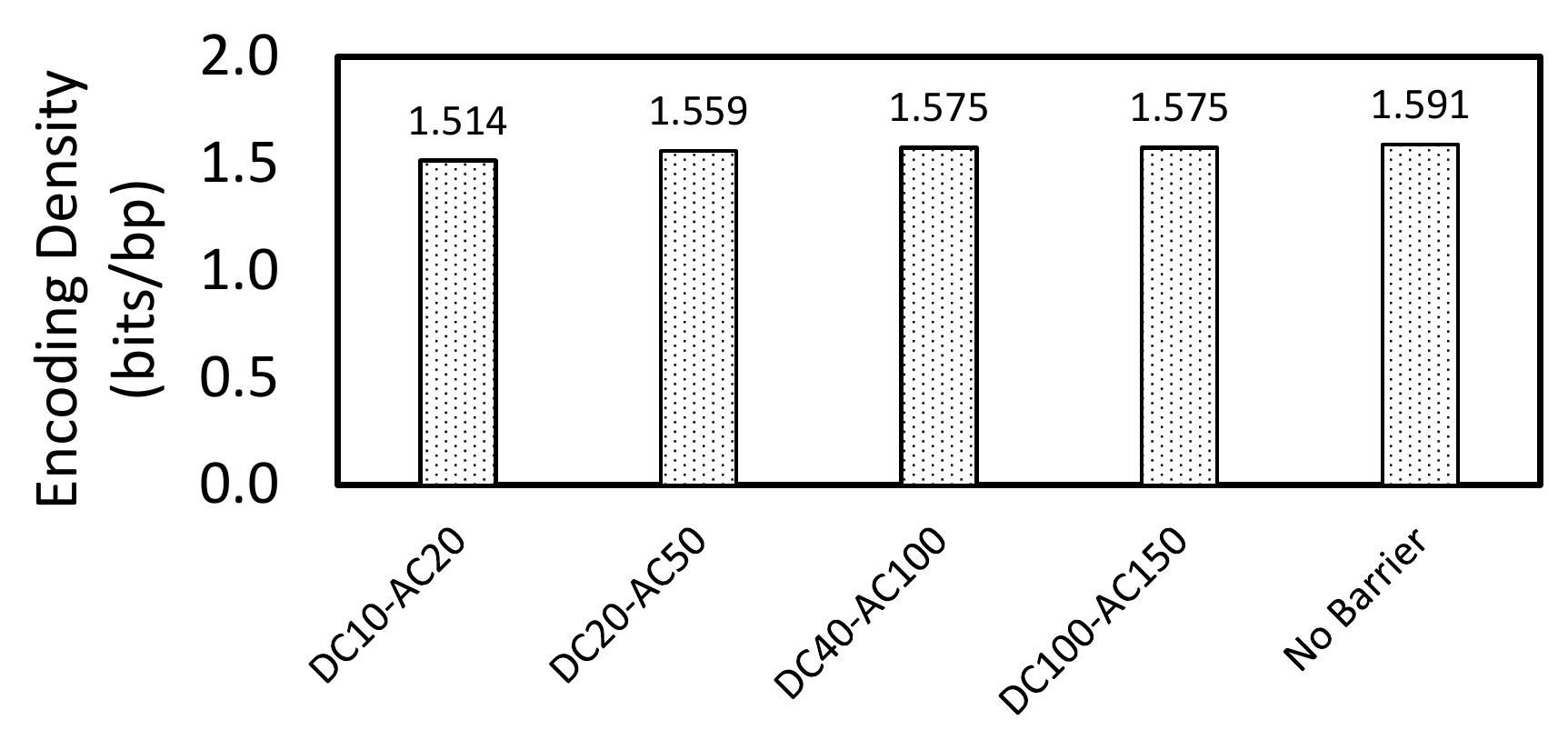}
	\caption{Capacity density with different schemes of adding barriers.}
	\label{fig:overhead}
\end{figure}

For the capacity overhead, IMG-DNA can introduce some capacity overhead due to adding extra barriers.   As shown in Figure~\ref{fig:overhead}, we investigate the effect of adding barriers on the encoding density with different partition lengths. From the figure, we can find that the encoding density is decreased from 1.591 bits/bp to 1.514 bits/bp when the PLs of DC and AC increase from 10 and 20 (i.e., DC10-AC20) to the maximum (i.e., No Barrier). Thus, the largest overhead is only about 0.07 bits/bp. Moreover, when we increase the PL to 100 bp for DC and 150 for AC (i.e., DC100-AC150), the encoding density can reach 1.575 bits/bp, which is only about 0.02 difference compared to the `No Barrier' case. In summary, there is a trade-off between encoding density and fault tolerance in the IMG-DNA scheme, and the overhead of the IMG-DNA is acceptable (about 1\% to 5\% capacity overhead).

\section{Related Work}\label{sec:related}
Two related research directions are DNA storage and approximate storage. The DNA storage~\cite{organick2018random, goldman2013towards, church2012next, blawat2016forward, bornholt2016dna} mainly focused on how to implement storing binary data in DNA storage. For example, Church et al.~\cite{church2012next} used the simple way to encode 1 and 0 by `A' or `C' and `T' or `G', respectively, which obtains an encoding density of 1 bit/nt. Goldman et al.~\cite{goldman2013towards} converted binary values to the trits by using Huffman code and then applied a rotating encoding to convert trits to DNA sequences. Thus, their studies did not consider using DNA storage as approximate storage. 

For the other type of research direction, researchers~\cite{guo2016high, yan2017customizing, kuo2019long, fan2019adaptive} applied fault-tolerant data such as images and videos in approximate storage devices based on the characteristics of the storage medium. For example, Kuo~et al.~\cite{kuo2019long} investigated the properties of JPEG-based images and SSDs and incorporated their properties together to improve the capacity of SSDs. Fan~\textit{et al.}~\cite{fan2019adaptive} proposed a new encoding scheme for DC and AC coefficients to lower the cost of using NAND flash. Those schemes based on traditional storage devices cannot be well applied to DNA storage since the unique properties of DNA storage systems should be exploited. Therefore, without considering the characteristics of DNA storage, their schemes are hard to obtain similar efficiency as they did in conventional storage systems.

\section{Conclusion}\label{sec:conclusion}
In this paper, we proposed a new scheme to efficiently store JPEG-based images in DNA storage, called IMG-DNA with a new DNA architecture by adding barriers to prevent error propagation in DNA storage. The barrier design is based on the encoding scheme and biochemical constraints in DNA storage. Then, we separately store DC and AC coefficients in different DNA sequences and use separate barrier schemes to further improve the error tolerance of DNA storage. Finally, the IMG-DNA scheme improves SSIM value by 2.6x - 19.7x and thus enhance the robustness of DNA storage compared to existing work.

\bibliographystyle{unsrt}
\bibliographystyle{ACM-Reference-Format}
\bibliography{acmart}	
\end{document}